\documentclass[%
reprint,pra,
superscriptaddress,
amsmath,amssymb,
aps,
]{revtex4}

\usepackage{graphicx}
\usepackage{dcolumn}
\usepackage{bm}
\usepackage{hyperref}
\usepackage[utf8]{inputenc}
\usepackage{subfigure}
\usepackage{cleveref}
\usepackage{amsmath}

\begin{document}
	
\preprint{APS/123-QED}

\title{Ghost translation}

\author{Wenhan Ren}
\affiliation{Institute for Quantum Science and Engineering, Texas A\&M University, College Station, Texas, 77843, USA}
\affiliation{School of Physics, Xi'an Jiaotong University, Xi'an, Shaanxi 710049, China}

\author{Xiaoyu Nie}
\affiliation{%
	Centre for Quantum Technologies, National University of Singapore, 3 Science Drive 2, Singapore 117543, Singapore}%

\author{Tao Peng}%
\email{taopeng@tamu.edu}
\affiliation{%
	Institute for Quantum Science and Engineering, Texas A\&M University, College Station, Texas, 77843, USA}%
\author{Marlan O. Scully}%
\email{scully@tamu.edu}
\affiliation{%
	Institute for Quantum Science and Engineering, Texas A\&M University, College Station, Texas, 77843, USA}%
\affiliation{%
Baylor University, Waco, 76706, USA}%
\affiliation{%
Princeton University, Princeton, NJ 08544, USA}%
\date{\today}

\begin{abstract}
Artificial intelligence has recently been widely used in computational imaging. The deep neural network (DNN) improves the signal-to-noise ratio of the retrieved images, whose quality is otherwise corrupted due to the low sampling ratio or noisy environments. This work proposes a new computational imaging scheme based on the sequence transduction mechanism with the transformer network. The simulation database assists the network in achieving signal translation ability. The experimental single-pixel detector's signal will be 'translated' into a 2D image in an end-to-end manner. High-quality images with no background noise can be retrieved at a sampling ratio as low as 2\%. The illumination patterns can be either well-designed speckle patterns for sub-Nyquist imaging or random speckle patterns. Moreover, our method is robust to noise interference. This translation mechanism opens a new direction for DNN-assisted ghost imaging and can be used in various computational imaging scenarios. 
\end{abstract}

	\maketitle
	
\section{Introduction}

Unlike conventional imaging, computational imaging (CI) utilizes both of the optical design and post-detection signal processing to create novel imaging systems. CI frequently outperforms the conventional approach, in terms of imaging resolution, speed, and the signal-to-noise ratio. It has been widely implemented in biophysics~\cite{cheng2021single}, tomography~\cite{peng2018micro,kingston2018ghost}, chemical microscopy~\cite{gattinger2019broadband,zeng2018hybrid}, and non-invasive imaging in complex media~\cite{stantchev2016noninvasive,zhao2022non}. One branch of CI is combining single-pixel detection technique, typically the computational ghost imaging (CGI)~\cite{shapiro2008computational,bromberg2009ghost}, with reconstruction algorithms to indirectly retrieve the object. The ghost image of an object is not directly obtained through a camera, but it is reconstructed through the spatial correlation of two light beams~\cite{Pittman1995Optical,valencia2005two,chen2009lensless}. In the computational fashion, specific speckle patterns are used for the spatial illumination~\cite{sun2017russian,zhang2017hadamard}. It can also be done on the detected signals in time domain~\cite{stockton2022tomographic, zhao2022imaging}, and compressive sensing (CS) on the sparsity properties~\cite{katz2009compressive, katkovnik2012compressive}. These techniques provide with advantages, {\it etc.}, high sampling efficiency~\cite{nie2022sub}, environmental noise robustness~\cite{nie2021noise}, high-resolution~\cite{cao2016resolution,bender2021circumventing}, broadband accessibility~\cite{pelliccia2016experimental,olivieri2020hyperspectral}, and non-optical availability~\cite{khakimov2016ghost,trimeche2020ion}. 
 
 With the prosperity and development of artificial intelligence in recent years, deep learning (DL) has been widely used for CI to achieve extremely low sampling ratios~\cite{wang2019learning,rizvi2020deepghost,wu2020deep}, or superresoltuion imaging~\cite{johnson2016perceptual,barbastathis2019use}. Deep neural networks (DNN) are used in the DL framework to improve the retrieved image quality after training using experimental~\cite{Lyu2017Deep} or simulated~\cite{song20220} data. Despite a variety of realizations, most DL imaging methods are based on a convolutional neural network(CNN). CNN is well-known for extracting the underlying features and visual structure~\cite{collobert2008unified}. The application of CNN to CI relies on using the kernel to extract local information over spatial neighborhoods of the imaging object. When the network is well-trained, it can extract the local features of the otherwise blurred image and give a clearer image with a higher signal-to-noise ratio. However, the underneath features of an object are not local, as suggested by the Fourier spectrum. Therefore, some essential features of the objects are ignored or less emphasized in the standard DL imaging schemes, especially in the sub-Nyquist condition, limiting their application to a broader range. 
 
  In this work, we propose and demonstrate a new CI algorithm based on the translation mechanism through the transformer neural network, namely, ghost translation (GT). Unlike the Recurrent Neural Network, in which the data is passed sequentially, The transformer framework processes the entire input all at once~\cite{vaswani2017attention}. It adopts the self-attention mechanism, deferentially weighting the significance of each part of the input data. The transformer has had great success in natural language processing~\cite{wolf2020transformers}. Most recently, it has also been demonstrated with excellent performance in the image recognition area~\cite{liu2021swin}. This work implements the translation of light intensity sequences into 2D images. Its unique attention mechanism enables each light intensity to pay attention to the information contained in other light intensities to enrich the connotation. Both simulations and experiments are demonstrated using two types of speckle patterns with different statistical properties. Results from the conventional CGI and CS methods using the same data sets are also presented for comparison. We also analyzed the resulting image quality with three typical evaluating indicators, and all the results suggest that GT outperforms the other methods to a great extent. Indeed, high-quality images can be retired through GT at a sampling ratio as low as 2\%. The framework also offers the ability to work in noisy environments.

\section{Method}\label{sec:method}

\subsection{theoretical approach}
\begin{figure}[htb!]
\centering
\includegraphics[width=0.6\linewidth]{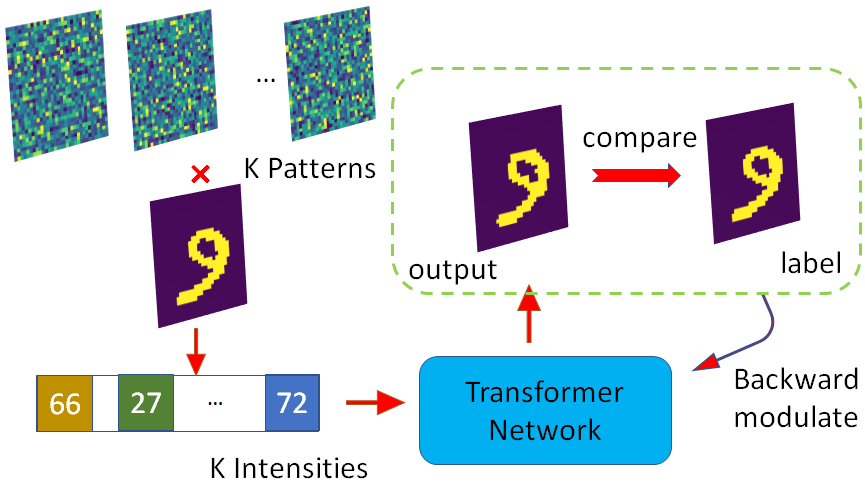}
\caption{Schematic of the ghost translation process. Bucket signals are applied to the ghost translation network and being ``translated'' into a 2D image.}
\label{fig:GT_principle}
\end{figure}
In a typical CI process, $K$ designed patterns $P_i$ (i = 1, 2, 3, $\cdots$, K) are illuminated onto the object $O (x, y)$ in sequence, the transmitted or reflected light intensities $I_i$ are then collected by a bucket detector:
\begin{equation}
 I_i = \int P_i(x,y)O(x,y)  dxdy.
\end{equation}

With the bucket signals,the GI can be reconstructed by the correlation function~\cite{chen2013100}
\begin{align}
O'(x,y) &=  \sum_{i=1}^{K} \langle \Delta I_i \Delta P_i(x,y) \rangle \cr
& \equiv \sum_{i=1}^{K} \langle I_i - \langle I_i\rangle \rangle \langle P_i(x,y) - \langle P_i(x,y)\rangle \rangle,
\end{align}

where $O'$(x,y) is the reconstructed object and $\langle \cdot \rangle$ denotes the ensemble average.

As shown in the above equation, although the bucket signal which contains the object information is only 1D, the 2D object can only be retrieved using a sequence of 2D speckle patterns. With CI techniques, a 2D object can be reconstructed from 1D data. The CS reconstruction algorithms search for the most sparse image in the compressible basis, it requires solving a convex optimization program, seeking for the image $\widetilde{O}_{cs}(x,y)$ which minimizes the $L_1$ norm in the sparse basis~\cite{katkovnik2012compressive}
\begin{align}
    arg~\text{min} || \Psi \{\widetilde{O}(x,y)\} ||_{L_1}, 
\end{align}
subject to 
\begin{align}
    \int P_i (x, y) \widetilde{O}(x,y) dx dy = S_i, \quad \forall_i = 1, 2, \cdots, M.
\end{align}
The problem of finding the image with the minimum $L_1$ norm can be solved efficiently~\cite{candes2006stable}. Deep learning has also been leveraged to solve the image CS reconstruction process~\cite{shi2017deep,zhang2018ista}. It worth noting that a recent work, an end-to-end DLCGI scheme was demonstrated to retrieve the 2D images from only 1D bucket signals~\cite{wang2019learning}. The process can be conceptually understood in two steps: the first is to Recovery of the illumination patterns $P_{i}(x, y)$, and the second step is to use the recovered patterns and the bucket signals for CGI processing, the results are then optimized by the DL network~\cite{jiao2020does}.

In our proposed scheme, we also use only the 1D bucket signals to reconstruct the 2D image. The 1D bucket is used as the input data, and the 2D image is the output. As shown in Fig.~\ref{fig:GT_principle}, the GT process can be represented as 
\begin{equation}
O'(x,y) = \mathcal{O} \{ I_i \},
\end{equation}
where $\mathcal{O}$ represents the ``translation'' process of the trained network. 
The cross-entropy loss is used to train the network and get the optimized solution~\cite{goodfellow2016deep}. The network in this training process can be written as \begin{equation}
O'  = L\{O(x,y),\mathcal{O}\{ I_i \}\}
\end{equation}

\subsection{Transformer network}
\begin{figure}[hbp!]
\centering
\includegraphics[width=0.6\linewidth]{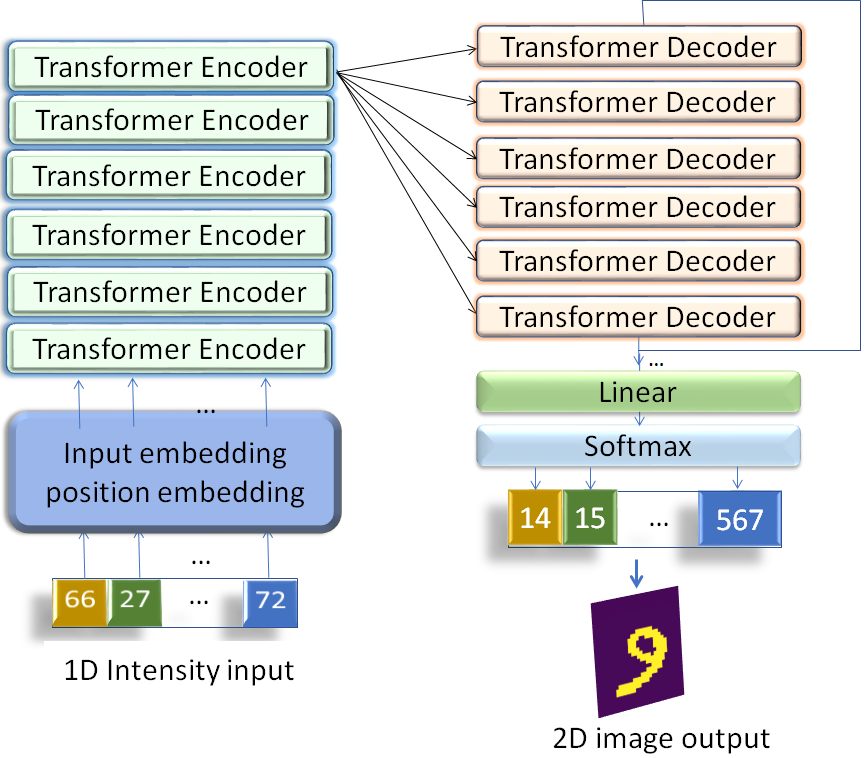}
\caption{The transformer network architecture for ghost translation.}  
\label{fig:GT_network}
\end{figure}
The transformer is a network based on the Seq2Seq model~\cite{sutskever2014sequence, cho2014learning}. As its name implies, Seq2Seq converts a sequence of tokens into another sequence of tokens. In general, the Seq2Seq model takes advantage of the Long-Short-Term-Memory(LSTM) model~\cite{schmidhuber1997long}, which can be used to give meaning to the sequence, make tokens in the sentence understand the positional relationship between them, and weaken or strengthen their significance. Seq2Seq is made up of encoders and  decoders. The encoders take input tokens and map them to high-dimensional spaces to create richer meanings, like a dictionary for words. The decoders read high-dimensional information generated by encoders and output it in various forms, such as words, characters, or expressions. 
The transformer is used most frequently in natural language processing because of these characteristics. Following this logic, our work will also treat the light intensity sequence and the two-dimensional image as two different languages. Through the transformer, we can customize the exclusive dictionary for these two languages, so any scene in the intensity sequence can be translated into a 2D image.
Our Transformer model, as shown in Fig.~\ref{fig:GT_network}, uses 6 encoder layers, 6 decoder layers. We firstly encode the input sequence to obtain rich embedding for each bucket signal. Each bucket signal value is modeled by a vector of size 512.  
The first layer of the encoder is the self-attention layer, which is an essential part of the encoder. It can detect the correlation between different light intensity vectors, no matter how far apart. The multi-head mechanism with eight attention heads is also used. These attention computations are then combined to produce the final attention score. This allows our model to discern relationships and nuances between different light intensities
Each decoder layer generates an output sequence by taking all the encoded information. The decoder and encoder layers have residual connections and layer normalization steps, along with feed-forward neural networks for additional processing. In this work, a total of  $N=32 \times 32$ pixel images are used, and each pixel is labeled such that the pixel number in the upper left corner is 1, and it is accumulated in turn, until the grid number in the lower right corner is 32$\times$32. Finally, all the grids that light up the numbers are left, resulting in a vector of outputs.

The proposed transformer network requires a training process based on a pre-prepared dataset, and after simulated training, it can directly reconstruct images from light intensity sequences. Here two training sets, i.e., the MNIST handwritten digits database and the Quick Draw smiley face dataset are used. All images are resized to 32$\times$32. During training, we set the batch to 32, train 32 images at a time, and set the maximum epoch to 100. We use the Adam optimizer with a learning rate of 0.001 and use learning rate warm up. We note here that, in the training process, the size of the input vector is fixed, and its size corresponds to the pixel multiples the sampling ratio, so the size of the output is not fixed. Training requires the output size of each batch to be the same, so we will first loop to get the maximum output bucket signal and complete the other outputs to this length. Use padding to 0 for completion, it will be ignored during training\cite{wang2019learning}.

\section{Experimental Results}
\subsection{Experimental setup}
\begin{figure}[hbp!]
\centering
\includegraphics[width=0.6\linewidth]{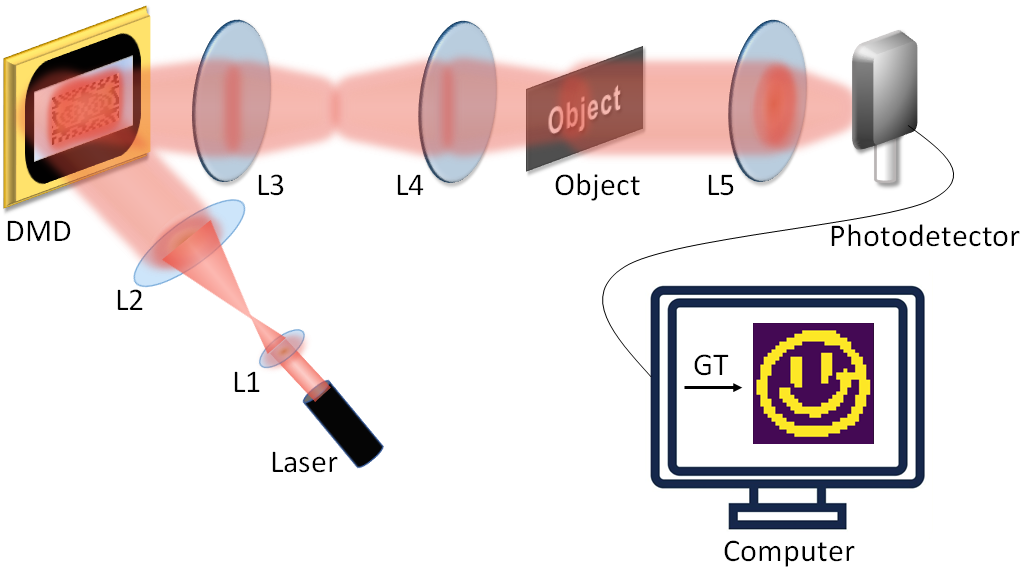}
\caption{Schematic setup. L1 and L2 consist of the laser beam expander. The expanded laser beam is modulated by the DMD with a sequence of displayed speckle patterns. The speckle pattern are then mapped onto the object plane through L3 and L4. The transmitted light is collected by a bucket detector which consists of a collection lens (L5) and a single-pixel photodetector. The intensity sequence is sent to the computer for the GT process.}  
\label{fig:setup}
\end{figure}
Our experimental setup is shown in Fig.~\ref{fig:setup}. As a typical CGI setup, the light source is a He-Ne CW laser with a wavelength of 633 $nm$. The laser is collimated and expanded by the lens system L1 and L2. The beam is then illuminated on the DMD (DLP4100), where the $K$ speckle patterns are displayed sequentially. In our experiment, the patterns are filled of  $32 \times 32$
independent pixels(each one counts $4\times 4$ DMD pixels). The object contains a total of 1024 pixels. The speckle patterns are then projected onto the object plane through a 4f system (L3 and l4). The transmitted light after the object is collected by lens L5 and detected by the single-pixel photodetector (Thorlabs PDA100A2). The single-pixel intensity signal can be used together with the patterns to generate CGI results. It can also be used for the CS algorithm, or for image retrieval though the GT network.

\begin{figure}[ht!]
\centering
\includegraphics[width=0.8\linewidth]{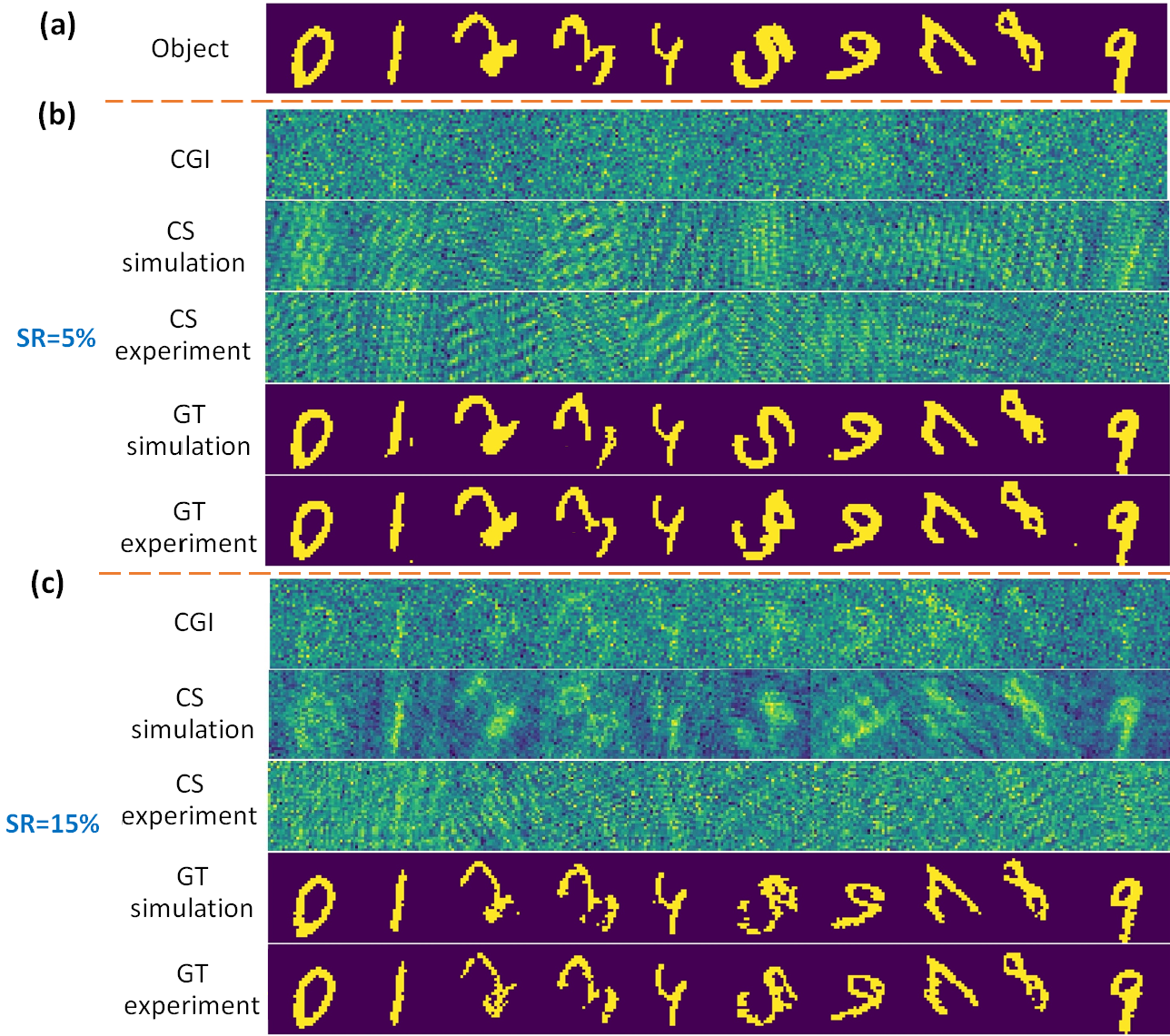}
\caption{GT results with Rayleigh speckles at sampling ratio of 5\% and 15\%. (a) The object, digits 0-9 from the MNIST handwritten dataset. (b) Results with SR=5\%. The first row is the traditional CGI. The second and third rows are the CS simulation and experimental results. The fourth and fifth are the GT simulation and experimental results. (c) Same as (b) with SR=15\%.}  
\label{fig:Rayleigh_results}
\end{figure}
\subsection{Results}

We used standard Rayleigh speckles with two different sampling ratio of SR=5\% (52 speckle patterns) and SR=15\% (154 speckle patterns) for the measurements. Ten objects of digits 0-9 from the MNIST handwritten digits database. We note here that these objects are chosen from the testing dataset, which is outside the training dataset. The results are shown in Fig.~\ref{fig:Rayleigh_results}. In Fig.~\ref{fig:Rayleigh_results} (b), we compare the traditional CGI simulation results, the CS simulation and experimental results, GT simulation the experimental GT results at SR=5\%. The simulation was done with no noise introduced. The experimental data was taken at a noise level of $\sim$ 9.6\% with a background light source illuminating the detector on purpose. Fig.~\ref{fig:Rayleigh_results} (c) presents similar results as of Fig.~\ref{fig:Rayleigh_results} (b) but with SR=15\%. For transitional CGI, it is only at 15\% sampling ratio when vague and noisy images are retrieve, the images are completely corrupted at the 5\% sampling ratio. For the CS algorithm, due to the small pixel sizes ($32\times 32$), therefore few patterns used, only simulation results at SR of 15\% can vague and noisy images be retrieved. When the noise is presented in the experimental data, no images can be retrieved at all. On the other hand, the GT can reconstruct high quality image even at the sampling ratio of 5\%, regardless the existence of noise. We notice that there is always no background in the retrieved images, which is another important feature of the GT scheme.  
\begin{figure}[pbt!]
\centering
\includegraphics[width=0.8\linewidth]{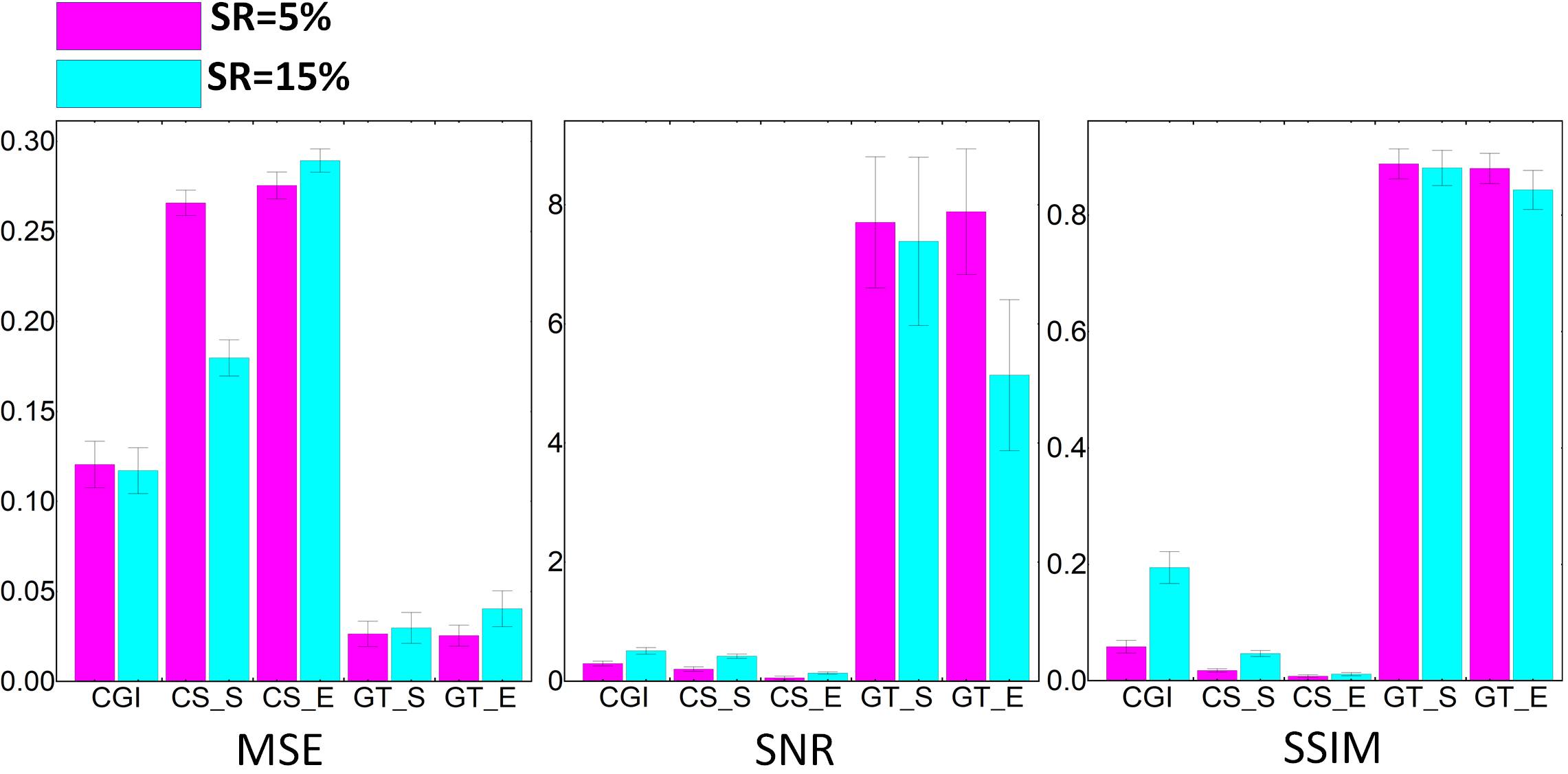}
\caption{MSE, SNR, and SSIM of the results shown in Fig.~\ref{fig:Rayleigh_results}. CS\_S, CS\_E, GT\_S, GT\_E are the CS simulation and experimental results, GT simulation and experimental results, respectively.}  
\label{fig:Rayleigh_SNR_SSIM}
\end{figure}
To quantitatively justify the quality of reconstructed images, we compare three evaluating indicators of image quality, \textit{i.e.}, the mean square error (MSE), signal-to-noise ratio (SNR), and the structural similarity index measure (SSIM):
\begin{align}
\mathrm{MSE} &= \frac{1}{N_{\mathrm{pixel}}}\sum_{n=1}^{N_{\mathrm{pixel}}}{(O'_i-O_i)^2},\cr
\mathrm{SNR} & =10\log{\frac{\sum_nO'_i}{\sum_n |O'_i-O_i|}},\cr
\mathrm{SSIM(O,O')} &= \frac{(2\bar{O}\bar{O'}+c_1)(2\sigma_{OO'}+c_2)}{(\bar{O}^2 +\bar{O'}^2+c_1)(\sigma^2_{O}+\sigma^2_{O'}+c_2)}.
\end{align}
Here $\bar{O}_n$ ($\bar{O'}_n$) is the intensity of the $n$-th pixel in the object (retrieved image), $\bar{O}$ ($\bar{O'}$) is the mean intensity of the pixels in the corresponding image, $\sigma_{O}^2$ ($\sigma^2_{O'}$) is the variance of the pixel intensity in the corresponding image,
$\sigma_{OO'}$ is the covariance of the pixel intensity in both images, $c_1$ and $c_2$ are regularization parameters. The MSE and SNR are standard parameters which characterize the image quality in terms of the signal and noise level, as compared to the original object, and SSIM describes changes in structural similarity between the objects, which is more sensitive to human perception~\cite{wang2004image}. These values are calculated with the ten digits from Fig.~\ref{fig:Rayleigh_results}. As shown in Fig.~\ref{fig:Rayleigh_SNR_SSIM}, the image quality of the CGI is poor at both sampling ratios, where 15\% is better than 5\%, as expected. Specifically, the SSIM at 15\% is much better as compared to the case of 5\%, since at 15\%, some of the digits can already be identified by eyes. 
Again, since the pixel size is small, the performance of CS is even worse than the traditional CGI. The image quality is further degraded when noise is introduced in the bucket signal.   
On the other hand, the GT results are always much better than that of the CGI and CS. We note here that, there is no significant difference between GT results with and without introduced noise. On the other hand, the GT results are always much better than that of the CGI and CS. We note here that there is no significant difference between GI results with and without introduced noise for the same sampling ratio. This is because the GT scheme uses attention to capture the global context information and ignores part of the noise interference. One may also notice that the 5\% sampling ratio results are slightly better than 15\%. This is partly because the image quality is already saturated at a 5\% sampling ratio. When only ten objects are randomly chosen for the compassion, the indicators might have better values at a 5\% sampling ratio. However, when the sampling ratio is 15\%, the detailed structure of the image can be better retrieved. This can be seen clearly in digits 3, 5, and 7, for instance. 
\begin{figure}[hpt!]
\centering
\includegraphics[width=0.6\linewidth]{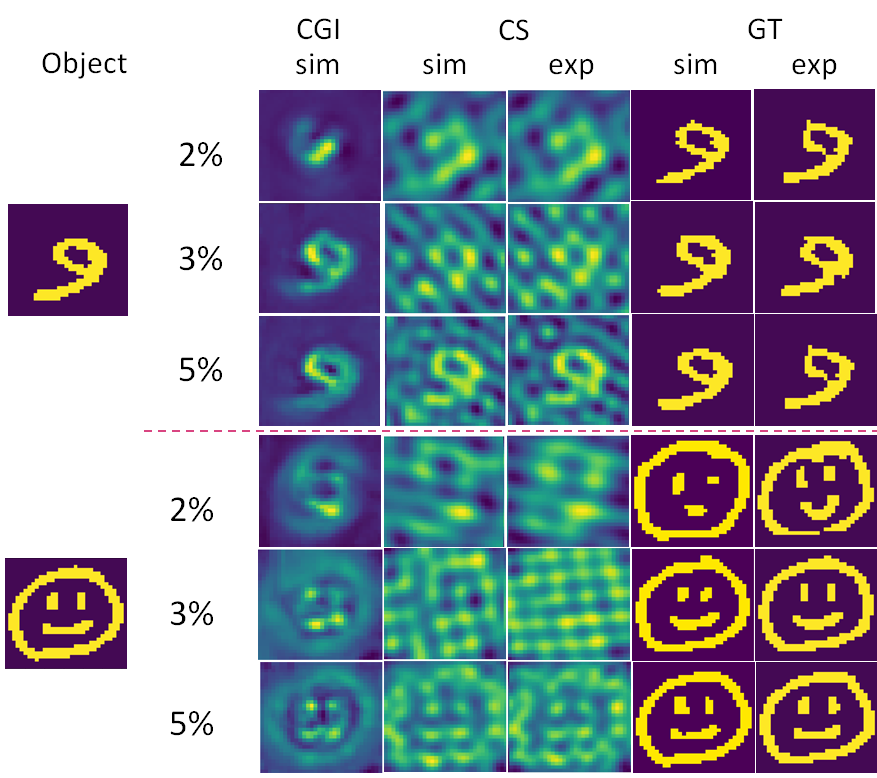}
\caption{Results of CGI, CS, and GT with customized speckles. The simulation (sim) and experimental (exp) results of one typical object from each dataset are presented at sampling ratios of 2\%, 3\%, and 5\%, respectively.}  
\label{fig:GT_DLpink}
\end{figure}
That being said, the most significant improvement when increasing the sampling ratio is the accuracy of the GT output, which is an essential metric that generally describes how the model performs~\cite{neimark2021video}. When 100 digits from the testing dataset are used, the accuracy of the GT is only $\sim~30\%$ at 5\% sampling ratio, but increases to 91\% at 15\% sampling ratio. In general, the accuracy can be further improved using more compute such as tokens, training dataset size, etc.

\subsection{GT with customized speckles}

\begin{figure*}[ht!]
\centering
\includegraphics[width=0.95\linewidth]{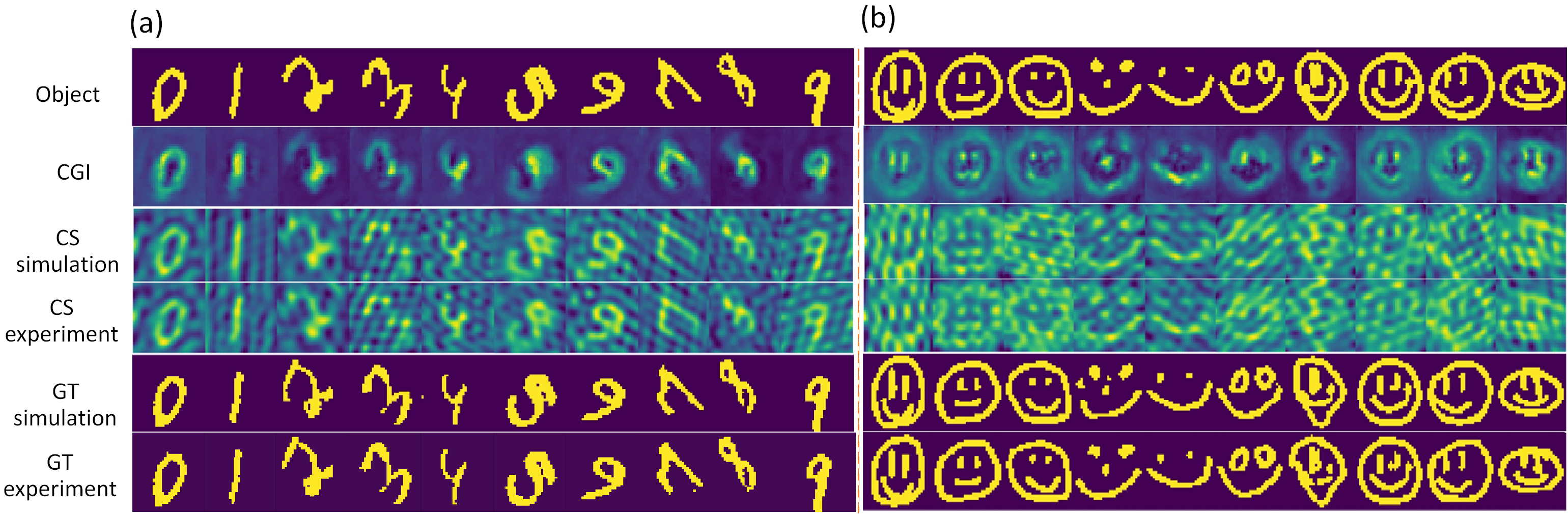}
\caption{Comparison of CGI, CS, and GT results of (a)digit numbers 0-9, and (b) ten randomly chosen smiley faces.  The sampling ratio of $\beta=5\%$ is used. The first column is the ground truth object. The second to sixth columns are the CGI simulation results, the CS simulation and experimental results, the GT simulation and experimental results, respectively.}  
\label{fig:GT_DLpink_005}
\end{figure*}

\begin{figure}[ht!]
\centering
\includegraphics[width=0.6\linewidth]{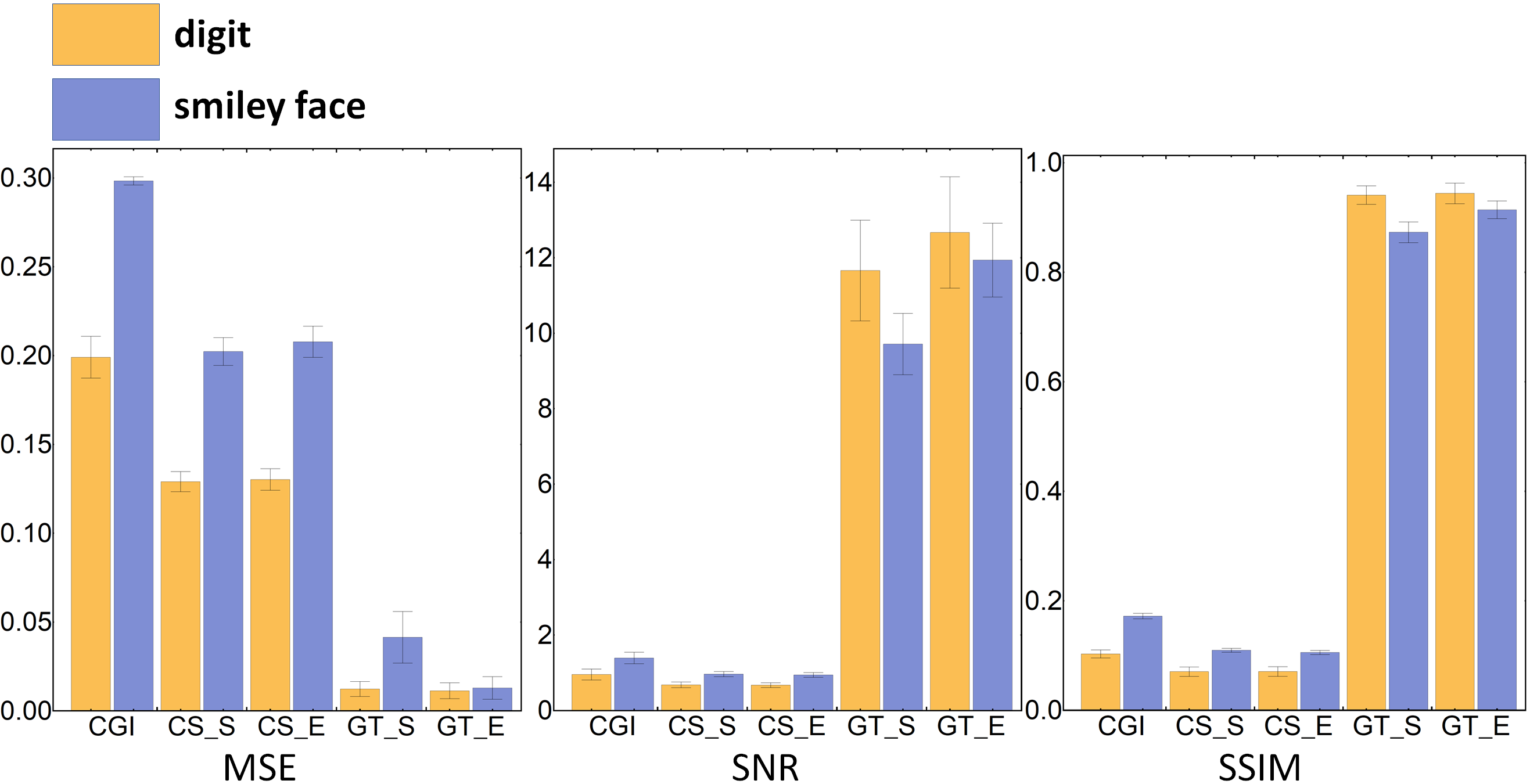}
\caption{MSE,SNR, and SSIM of the results shown in Fig.~\ref{fig:GT_DLpink_005}.}
\label{fig:DLpink_MSE_SNR_SSIM}
\end{figure}

Next, we show that the GT results can be further improved when customized speckles are used. When standard random speckle patterns interact with the object, the intrinsic connection or correlation between the bucket signals is from the object itself. Our recent work has demonstrated that speckles can be optimized via deep learning at the sub-Nyquist condition~\cite{nie2021deep}. Due to the unique statistic feature, the results outperform the deep learning GI process when the speckles are used for the CGI. The particular statistics of the speckles also suggest that the bucket signals have a unique connection. This type of connection could benefit the attention mechanism used in the GT network. We, therefore, use these speckle patterns for the GT process for better performance. Here we used three different sampling ratios at 2\% (21 speckle pattern), 3\% (31 speckle patterns), and 5\%. In addition to the MNIST dataset, we used the Quick Draw smiley face dataset to examine the generalization of the GT method. The simulation and experimental results of one typical example each from the two datasets are shown in Fig.~\ref{fig:GT_DLpink}. It is clearly seen that the GT results can tentatively retrieve both images at the sampling ratio as low as 2\%. For sampling ratios of 2\%, 3\%, and 5\%, the overall accuracy is 36\%, 66\%, and 80\%, respectively. We notice that at 5\%, the GT results are almost identical to the original objects. 

More evidence is shown in Fig.~\ref{fig:GT_DLpink_005}, where ten digits and ten randomly selected smiley faces are used. It can be seen that almost all the ten digits and ten smiley faces are retrieved in GT simulation and experiments. Not only there is no background noise in the retrieved images, but the detailed structures are almost recovered from the network. This is a significant difference than when the Rayleigh speckle patterns are used to generate the bucket signals. This could be mainly due to that, the self-attention mechanism highly emphasizes the connection between the encoded inputs. When the input bucket signal are generated with speckle patterns that are random (therefore almost orthogonal to each other), less intrinsic connections are encoded in the bucket signals. On the other hand, due to the unique property of the customized speckle patterns, the bucket signals also contain enough information that can later be recognized by the transformer network. Again, even when noise is introduced in the experiment, the results are unaffected. The results of the other two computational imaging methods, CGI and CS, are also presented in Fig.~\ref{fig:GT_DLpink_005}. The background noise is suppressed in general, when customized speckle patterns are used. However, the overall imaging quality is still far worse than the GT results.

The results described by the image quality indicators are depicted in Fig.~\ref{fig:DLpink_MSE_SNR_SSIM}. The MSE, SNR, and SSIM values all suggest that the quality of the digits are better than the smiley faces. This could be the fact that the Quick Draw dataset is more diverse in the structure since it is created casually. However, as indicated by the MSE and SSIM values, the experimental results for both datasets are almost identical to the original objects. Further improvement, especially in regard to the accuracy,  may be achieved by increasing the training set size, imaging size, increasing the sampling ratio, or optimising the network, etc.

\section{Conclusion}
In summary, we have developed and demonstrated a translation mechanism based computational imaging. The object image can be translated only from the bucket detector signals. We have shown that the transformer can be well trained using only simulation data, so the cost of training can be significantly reduced. The proposed method was verified on two different datasets at different sampling ratios through simulations and experiments. Our observation suggests that the proposed method can restore the object image at extremely low sampling ratios. It is also robust to noise interference due to the unique attention mechanism used in the network. This has significant potential to increase the time efficiency of data acquisition in practical applications, especially in noisy environments.

\noindent\textbf{Funding.}
Air Force Office of Scientific Research (Award No. FA9550-20-1-0366 DEF), Office of Naval Research (Award No. N00014-20-1-2184), Robert A. Welch Foundation (Grant No. A-1261), National Science Foundation (Grant No. PHY-2013771).

\noindent\textbf{Disclosures.} The authors declare no conflicts of interest.


\begin{thebibliography}{49}
	\expandafter\ifx\csname natexlab\endcsname\relax\def\natexlab#1{#1}\fi
	\expandafter\ifx\csname bibnamefont\endcsname\relax
	\def\bibnamefont#1{#1}\fi
	\expandafter\ifx\csname bibfnamefont\endcsname\relax
	\def\bibfnamefont#1{#1}\fi
	\expandafter\ifx\csname citenamefont\endcsname\relax
	\def\citenamefont#1{#1}\fi
	\expandafter\ifx\csname url\endcsname\relax
	\def\url#1{\texttt{#1}}\fi
	\expandafter\ifx\csname urlprefix\endcsname\relax\def\urlprefix{URL }\fi
	\providecommand{\bibinfo}[2]{#2}
	\providecommand{\eprint}[2][]{\url{#2}}
	
	\bibitem[{\citenamefont{Cheng et~al.}(2021)\citenamefont{Cheng, Fu, Kim, Song,
			Li, Xue, Yi, and Tian}}]{cheng2021single}
	\bibinfo{author}{\bibfnamefont{S.}~\bibnamefont{Cheng}},
	\bibinfo{author}{\bibfnamefont{S.}~\bibnamefont{Fu}},
	\bibinfo{author}{\bibfnamefont{Y.~M.} \bibnamefont{Kim}},
	\bibinfo{author}{\bibfnamefont{W.}~\bibnamefont{Song}},
	\bibinfo{author}{\bibfnamefont{Y.}~\bibnamefont{Li}},
	\bibinfo{author}{\bibfnamefont{Y.}~\bibnamefont{Xue}},
	\bibinfo{author}{\bibfnamefont{J.}~\bibnamefont{Yi}}, \bibnamefont{and}
	\bibinfo{author}{\bibfnamefont{L.}~\bibnamefont{Tian}},
	\bibinfo{journal}{Science advances} \textbf{\bibinfo{volume}{7}},
	\bibinfo{pages}{eabe0431} (\bibinfo{year}{2021}).
	
	\bibitem[{\citenamefont{Peng et~al.}(2018)\citenamefont{Peng, Yao, Cheng,
			Zhang, Li, Zheng, and Zhong}}]{peng2018micro}
	\bibinfo{author}{\bibfnamefont{J.}~\bibnamefont{Peng}},
	\bibinfo{author}{\bibfnamefont{M.}~\bibnamefont{Yao}},
	\bibinfo{author}{\bibfnamefont{J.}~\bibnamefont{Cheng}},
	\bibinfo{author}{\bibfnamefont{Z.}~\bibnamefont{Zhang}},
	\bibinfo{author}{\bibfnamefont{S.}~\bibnamefont{Li}},
	\bibinfo{author}{\bibfnamefont{G.}~\bibnamefont{Zheng}}, \bibnamefont{and}
	\bibinfo{author}{\bibfnamefont{J.}~\bibnamefont{Zhong}},
	\bibinfo{journal}{Optics express} \textbf{\bibinfo{volume}{26}},
	\bibinfo{pages}{31094} (\bibinfo{year}{2018}).
	
	\bibitem[{\citenamefont{Kingston et~al.}(2018)\citenamefont{Kingston,
			Pelliccia, Rack, Olbinado, Cheng, Myers, and Paganin}}]{kingston2018ghost}
	\bibinfo{author}{\bibfnamefont{A.~M.} \bibnamefont{Kingston}},
	\bibinfo{author}{\bibfnamefont{D.}~\bibnamefont{Pelliccia}},
	\bibinfo{author}{\bibfnamefont{A.}~\bibnamefont{Rack}},
	\bibinfo{author}{\bibfnamefont{M.~P.} \bibnamefont{Olbinado}},
	\bibinfo{author}{\bibfnamefont{Y.}~\bibnamefont{Cheng}},
	\bibinfo{author}{\bibfnamefont{G.~R.} \bibnamefont{Myers}}, \bibnamefont{and}
	\bibinfo{author}{\bibfnamefont{D.~M.} \bibnamefont{Paganin}},
	\bibinfo{journal}{Optica} \textbf{\bibinfo{volume}{5}}, \bibinfo{pages}{1516}
	(\bibinfo{year}{2018}).
	
	\bibitem[{\citenamefont{Gattinger et~al.}(2019)\citenamefont{Gattinger, Kilgus,
			Zorin, Langer, Nikzad-Langerodi, Rankl, Gr{\"o}schl, and
			Brandstetter}}]{gattinger2019broadband}
	\bibinfo{author}{\bibfnamefont{P.}~\bibnamefont{Gattinger}},
	\bibinfo{author}{\bibfnamefont{J.}~\bibnamefont{Kilgus}},
	\bibinfo{author}{\bibfnamefont{I.}~\bibnamefont{Zorin}},
	\bibinfo{author}{\bibfnamefont{G.}~\bibnamefont{Langer}},
	\bibinfo{author}{\bibfnamefont{R.}~\bibnamefont{Nikzad-Langerodi}},
	\bibinfo{author}{\bibfnamefont{C.}~\bibnamefont{Rankl}},
	\bibinfo{author}{\bibfnamefont{M.}~\bibnamefont{Gr{\"o}schl}},
	\bibnamefont{and}
	\bibinfo{author}{\bibfnamefont{M.}~\bibnamefont{Brandstetter}},
	\bibinfo{journal}{Optics express} \textbf{\bibinfo{volume}{27}},
	\bibinfo{pages}{12666} (\bibinfo{year}{2019}).
	
	\bibitem[{\citenamefont{Zeng et~al.}(2018)\citenamefont{Zeng, Huang, Singh,
			Yao, Azad, Mohite, Taylor, Smith, and Chen}}]{zeng2018hybrid}
	\bibinfo{author}{\bibfnamefont{B.}~\bibnamefont{Zeng}},
	\bibinfo{author}{\bibfnamefont{Z.}~\bibnamefont{Huang}},
	\bibinfo{author}{\bibfnamefont{A.}~\bibnamefont{Singh}},
	\bibinfo{author}{\bibfnamefont{Y.}~\bibnamefont{Yao}},
	\bibinfo{author}{\bibfnamefont{A.~K.} \bibnamefont{Azad}},
	\bibinfo{author}{\bibfnamefont{A.~D.} \bibnamefont{Mohite}},
	\bibinfo{author}{\bibfnamefont{A.~J.} \bibnamefont{Taylor}},
	\bibinfo{author}{\bibfnamefont{D.~R.} \bibnamefont{Smith}}, \bibnamefont{and}
	\bibinfo{author}{\bibfnamefont{H.-T.} \bibnamefont{Chen}},
	\bibinfo{journal}{Light: Science \& Applications}
	\textbf{\bibinfo{volume}{7}}, \bibinfo{pages}{1} (\bibinfo{year}{2018}).
	
	\bibitem[{\citenamefont{Stantchev et~al.}(2016)\citenamefont{Stantchev, Sun,
			Hornett, Hobson, Gibson, Padgett, and Hendry}}]{stantchev2016noninvasive}
	\bibinfo{author}{\bibfnamefont{R.~I.} \bibnamefont{Stantchev}},
	\bibinfo{author}{\bibfnamefont{B.}~\bibnamefont{Sun}},
	\bibinfo{author}{\bibfnamefont{S.~M.} \bibnamefont{Hornett}},
	\bibinfo{author}{\bibfnamefont{P.~A.} \bibnamefont{Hobson}},
	\bibinfo{author}{\bibfnamefont{G.~M.} \bibnamefont{Gibson}},
	\bibinfo{author}{\bibfnamefont{M.~J.} \bibnamefont{Padgett}},
	\bibnamefont{and} \bibinfo{author}{\bibfnamefont{E.}~\bibnamefont{Hendry}},
	\bibinfo{journal}{Science advances} \textbf{\bibinfo{volume}{2}},
	\bibinfo{pages}{e1600190} (\bibinfo{year}{2016}).
	
	\bibitem[{\citenamefont{Zhao et~al.}(2022{\natexlab{a}})\citenamefont{Zhao,
			Peng, Zhang, Zubairy, Shih, and Scully}}]{zhao2022non}
	\bibinfo{author}{\bibfnamefont{X.}~\bibnamefont{Zhao}},
	\bibinfo{author}{\bibfnamefont{T.}~\bibnamefont{Peng}},
	\bibinfo{author}{\bibfnamefont{L.}~\bibnamefont{Zhang}},
	\bibinfo{author}{\bibfnamefont{M.~S.} \bibnamefont{Zubairy}},
	\bibinfo{author}{\bibfnamefont{Y.}~\bibnamefont{Shih}}, \bibnamefont{and}
	\bibinfo{author}{\bibfnamefont{M.~O.} \bibnamefont{Scully}},
	\bibinfo{journal}{IEEE Photonics Journal}
	(\bibinfo{year}{2022}{\natexlab{a}}).
	
	\bibitem[{\citenamefont{Shapiro}(2008)}]{shapiro2008computational}
	\bibinfo{author}{\bibfnamefont{J.~H.} \bibnamefont{Shapiro}},
	\bibinfo{journal}{Physical Review A} \textbf{\bibinfo{volume}{78}},
	\bibinfo{pages}{061802} (\bibinfo{year}{2008}).
	
	\bibitem[{\citenamefont{Bromberg et~al.}(2009)\citenamefont{Bromberg, Katz, and
			Silberberg}}]{bromberg2009ghost}
	\bibinfo{author}{\bibfnamefont{Y.}~\bibnamefont{Bromberg}},
	\bibinfo{author}{\bibfnamefont{O.}~\bibnamefont{Katz}}, \bibnamefont{and}
	\bibinfo{author}{\bibfnamefont{Y.}~\bibnamefont{Silberberg}},
	\bibinfo{journal}{Physical Review A} \textbf{\bibinfo{volume}{79}},
	\bibinfo{pages}{053840} (\bibinfo{year}{2009}).
	
	\bibitem[{\citenamefont{Pittman et~al.}(1995)\citenamefont{Pittman, Shih,
			Strekalov, and Sergienko}}]{Pittman1995Optical}
	\bibinfo{author}{\bibfnamefont{T.~B.} \bibnamefont{Pittman}},
	\bibinfo{author}{\bibfnamefont{Y.~H.} \bibnamefont{Shih}},
	\bibinfo{author}{\bibfnamefont{D.~V.} \bibnamefont{Strekalov}},
	\bibnamefont{and} \bibinfo{author}{\bibfnamefont{A.~V.}
		\bibnamefont{Sergienko}}, \bibinfo{journal}{Physical Review A}
	\textbf{\bibinfo{volume}{52}}, \bibinfo{pages}{R3429} (\bibinfo{year}{1995}).
	
	\bibitem[{\citenamefont{Valencia et~al.}(2005)\citenamefont{Valencia,
			Scarcelli, D’Angelo, and Shih}}]{valencia2005two}
	\bibinfo{author}{\bibfnamefont{A.}~\bibnamefont{Valencia}},
	\bibinfo{author}{\bibfnamefont{G.}~\bibnamefont{Scarcelli}},
	\bibinfo{author}{\bibfnamefont{M.}~\bibnamefont{D’Angelo}},
	\bibnamefont{and} \bibinfo{author}{\bibfnamefont{Y.}~\bibnamefont{Shih}},
	\bibinfo{journal}{Physical Review Letters} \textbf{\bibinfo{volume}{94}},
	\bibinfo{pages}{063601} (\bibinfo{year}{2005}).
	
	\bibitem[{\citenamefont{Chen et~al.}(2009)\citenamefont{Chen, Liu, Luo, and
			Wu}}]{chen2009lensless}
	\bibinfo{author}{\bibfnamefont{X.-H.} \bibnamefont{Chen}},
	\bibinfo{author}{\bibfnamefont{Q.}~\bibnamefont{Liu}},
	\bibinfo{author}{\bibfnamefont{K.-H.} \bibnamefont{Luo}}, \bibnamefont{and}
	\bibinfo{author}{\bibfnamefont{L.-A.} \bibnamefont{Wu}},
	\bibinfo{journal}{Optics Letters} \textbf{\bibinfo{volume}{34}},
	\bibinfo{pages}{695} (\bibinfo{year}{2009}).
	
	\bibitem[{\citenamefont{Sun et~al.}(2017)\citenamefont{Sun, Meng, Edgar,
			Padgett, and Radwell}}]{sun2017russian}
	\bibinfo{author}{\bibfnamefont{M.-J.} \bibnamefont{Sun}},
	\bibinfo{author}{\bibfnamefont{L.-T.} \bibnamefont{Meng}},
	\bibinfo{author}{\bibfnamefont{M.~P.} \bibnamefont{Edgar}},
	\bibinfo{author}{\bibfnamefont{M.~J.} \bibnamefont{Padgett}},
	\bibnamefont{and} \bibinfo{author}{\bibfnamefont{N.}~\bibnamefont{Radwell}},
	\bibinfo{journal}{Scientific Reports} \textbf{\bibinfo{volume}{7}},
	\bibinfo{pages}{1} (\bibinfo{year}{2017}).
	
	\bibitem[{\citenamefont{Zhang et~al.}(2017)\citenamefont{Zhang, Wang, Zheng,
			and Zhong}}]{zhang2017hadamard}
	\bibinfo{author}{\bibfnamefont{Z.}~\bibnamefont{Zhang}},
	\bibinfo{author}{\bibfnamefont{X.}~\bibnamefont{Wang}},
	\bibinfo{author}{\bibfnamefont{G.}~\bibnamefont{Zheng}}, \bibnamefont{and}
	\bibinfo{author}{\bibfnamefont{J.}~\bibnamefont{Zhong}},
	\bibinfo{journal}{Optics Express} \textbf{\bibinfo{volume}{25}},
	\bibinfo{pages}{19619} (\bibinfo{year}{2017}).
	
	\bibitem[{\citenamefont{Stockton et~al.}(2022)\citenamefont{Stockton, Murray,
			Field, Squier, Pezeshki, and Bartels}}]{stockton2022tomographic}
	\bibinfo{author}{\bibfnamefont{P.}~\bibnamefont{Stockton}},
	\bibinfo{author}{\bibfnamefont{G.}~\bibnamefont{Murray}},
	\bibinfo{author}{\bibfnamefont{J.~J.} \bibnamefont{Field}},
	\bibinfo{author}{\bibfnamefont{J.}~\bibnamefont{Squier}},
	\bibinfo{author}{\bibfnamefont{A.}~\bibnamefont{Pezeshki}}, \bibnamefont{and}
	\bibinfo{author}{\bibfnamefont{R.~A.} \bibnamefont{Bartels}},
	\bibinfo{journal}{Optics Communications} p. \bibinfo{pages}{128401}
	(\bibinfo{year}{2022}).
	
	\bibitem[{\citenamefont{Zhao et~al.}(2022{\natexlab{b}})\citenamefont{Zhao,
			Nie, Yi, Peng, and Scully}}]{zhao2022imaging}
	\bibinfo{author}{\bibfnamefont{X.}~\bibnamefont{Zhao}},
	\bibinfo{author}{\bibfnamefont{X.}~\bibnamefont{Nie}},
	\bibinfo{author}{\bibfnamefont{Z.}~\bibnamefont{Yi}},
	\bibinfo{author}{\bibfnamefont{T.}~\bibnamefont{Peng}}, \bibnamefont{and}
	\bibinfo{author}{\bibfnamefont{M.~O.} \bibnamefont{Scully}},
	\bibinfo{journal}{Photonics Research} \textbf{\bibinfo{volume}{10}},
	\bibinfo{pages}{1689} (\bibinfo{year}{2022}{\natexlab{b}}).
	
	\bibitem[{\citenamefont{Katz et~al.}(2009)\citenamefont{Katz, Bromberg, and
			Silberberg}}]{katz2009compressive}
	\bibinfo{author}{\bibfnamefont{O.}~\bibnamefont{Katz}},
	\bibinfo{author}{\bibfnamefont{Y.}~\bibnamefont{Bromberg}}, \bibnamefont{and}
	\bibinfo{author}{\bibfnamefont{Y.}~\bibnamefont{Silberberg}},
	\bibinfo{journal}{Applied Physics Letters} \textbf{\bibinfo{volume}{95}},
	\bibinfo{pages}{739} (\bibinfo{year}{2009}).
	
	\bibitem[{\citenamefont{Katkovnik and Astola}(2012)}]{katkovnik2012compressive}
	\bibinfo{author}{\bibfnamefont{V.}~\bibnamefont{Katkovnik}} \bibnamefont{and}
	\bibinfo{author}{\bibfnamefont{J.}~\bibnamefont{Astola}},
	\bibinfo{journal}{JOSA A} \textbf{\bibinfo{volume}{29}},
	\bibinfo{pages}{1556} (\bibinfo{year}{2012}).
	
	\bibitem[{\citenamefont{Nie et~al.}(2022)\citenamefont{Nie, Zhao, Peng, and
			Scully}}]{nie2022sub}
	\bibinfo{author}{\bibfnamefont{X.}~\bibnamefont{Nie}},
	\bibinfo{author}{\bibfnamefont{X.}~\bibnamefont{Zhao}},
	\bibinfo{author}{\bibfnamefont{T.}~\bibnamefont{Peng}}, \bibnamefont{and}
	\bibinfo{author}{\bibfnamefont{M.~O.} \bibnamefont{Scully}},
	\bibinfo{journal}{Physical Review A} \textbf{\bibinfo{volume}{105}},
	\bibinfo{pages}{043525} (\bibinfo{year}{2022}).
	
	\bibitem[{\citenamefont{Nie et~al.}(2021{\natexlab{a}})\citenamefont{Nie, Yang,
			Liu, Zhao, Nessler, Peng, Zubairy, and Scully}}]{nie2021noise}
	\bibinfo{author}{\bibfnamefont{X.}~\bibnamefont{Nie}},
	\bibinfo{author}{\bibfnamefont{F.}~\bibnamefont{Yang}},
	\bibinfo{author}{\bibfnamefont{X.}~\bibnamefont{Liu}},
	\bibinfo{author}{\bibfnamefont{X.}~\bibnamefont{Zhao}},
	\bibinfo{author}{\bibfnamefont{R.}~\bibnamefont{Nessler}},
	\bibinfo{author}{\bibfnamefont{T.}~\bibnamefont{Peng}},
	\bibinfo{author}{\bibfnamefont{M.~S.} \bibnamefont{Zubairy}},
	\bibnamefont{and} \bibinfo{author}{\bibfnamefont{M.~O.}
		\bibnamefont{Scully}}, \bibinfo{journal}{Physical Review A}
	\textbf{\bibinfo{volume}{104}}, \bibinfo{pages}{013513}
	(\bibinfo{year}{2021}{\natexlab{a}}).
	
	\bibitem[{\citenamefont{Cao et~al.}(2016)\citenamefont{Cao, Yang, Wang, Qiu,
			Wei, Gao, and Li}}]{cao2016resolution}
	\bibinfo{author}{\bibfnamefont{M.}~\bibnamefont{Cao}},
	\bibinfo{author}{\bibfnamefont{X.}~\bibnamefont{Yang}},
	\bibinfo{author}{\bibfnamefont{J.}~\bibnamefont{Wang}},
	\bibinfo{author}{\bibfnamefont{S.}~\bibnamefont{Qiu}},
	\bibinfo{author}{\bibfnamefont{D.}~\bibnamefont{Wei}},
	\bibinfo{author}{\bibfnamefont{H.}~\bibnamefont{Gao}}, \bibnamefont{and}
	\bibinfo{author}{\bibfnamefont{F.}~\bibnamefont{Li}},
	\bibinfo{journal}{Optics letters} \textbf{\bibinfo{volume}{41}},
	\bibinfo{pages}{5349} (\bibinfo{year}{2016}).
	
	\bibitem[{\citenamefont{Bender et~al.}(2021)\citenamefont{Bender, Sun,
			Y{\i}lmaz, Bewersdorf, and Cao}}]{bender2021circumventing}
	\bibinfo{author}{\bibfnamefont{N.}~\bibnamefont{Bender}},
	\bibinfo{author}{\bibfnamefont{M.}~\bibnamefont{Sun}},
	\bibinfo{author}{\bibfnamefont{H.}~\bibnamefont{Y{\i}lmaz}},
	\bibinfo{author}{\bibfnamefont{J.}~\bibnamefont{Bewersdorf}},
	\bibnamefont{and} \bibinfo{author}{\bibfnamefont{H.}~\bibnamefont{Cao}},
	\bibinfo{journal}{Optica} \textbf{\bibinfo{volume}{8}}, \bibinfo{pages}{122}
	(\bibinfo{year}{2021}).
	
	\bibitem[{\citenamefont{Pelliccia et~al.}(2016)\citenamefont{Pelliccia, Rack,
			Scheel, Cantelli, and Paganin}}]{pelliccia2016experimental}
	\bibinfo{author}{\bibfnamefont{D.}~\bibnamefont{Pelliccia}},
	\bibinfo{author}{\bibfnamefont{A.}~\bibnamefont{Rack}},
	\bibinfo{author}{\bibfnamefont{M.}~\bibnamefont{Scheel}},
	\bibinfo{author}{\bibfnamefont{V.}~\bibnamefont{Cantelli}}, \bibnamefont{and}
	\bibinfo{author}{\bibfnamefont{D.~M.} \bibnamefont{Paganin}},
	\bibinfo{journal}{Physical review letters} \textbf{\bibinfo{volume}{117}},
	\bibinfo{pages}{113902} (\bibinfo{year}{2016}).
	
	\bibitem[{\citenamefont{Olivieri et~al.}(2020)\citenamefont{Olivieri, Gongora,
			Peters, Cecconi, Cutrona, Tunesi, Tucker, Pasquazi, and
			Peccianti}}]{olivieri2020hyperspectral}
	\bibinfo{author}{\bibfnamefont{L.}~\bibnamefont{Olivieri}},
	\bibinfo{author}{\bibfnamefont{J.~S.~T.} \bibnamefont{Gongora}},
	\bibinfo{author}{\bibfnamefont{L.}~\bibnamefont{Peters}},
	\bibinfo{author}{\bibfnamefont{V.}~\bibnamefont{Cecconi}},
	\bibinfo{author}{\bibfnamefont{A.}~\bibnamefont{Cutrona}},
	\bibinfo{author}{\bibfnamefont{J.}~\bibnamefont{Tunesi}},
	\bibinfo{author}{\bibfnamefont{R.}~\bibnamefont{Tucker}},
	\bibinfo{author}{\bibfnamefont{A.}~\bibnamefont{Pasquazi}}, \bibnamefont{and}
	\bibinfo{author}{\bibfnamefont{M.}~\bibnamefont{Peccianti}},
	\bibinfo{journal}{Optica} \textbf{\bibinfo{volume}{7}}, \bibinfo{pages}{186}
	(\bibinfo{year}{2020}).
	
	\bibitem[{\citenamefont{Khakimov et~al.}(2016)\citenamefont{Khakimov, Henson,
			Shin, Hodgman, Dall, Baldwin, and Truscott}}]{khakimov2016ghost}
	\bibinfo{author}{\bibfnamefont{R.~I.} \bibnamefont{Khakimov}},
	\bibinfo{author}{\bibfnamefont{B.}~\bibnamefont{Henson}},
	\bibinfo{author}{\bibfnamefont{D.}~\bibnamefont{Shin}},
	\bibinfo{author}{\bibfnamefont{S.}~\bibnamefont{Hodgman}},
	\bibinfo{author}{\bibfnamefont{R.}~\bibnamefont{Dall}},
	\bibinfo{author}{\bibfnamefont{K.}~\bibnamefont{Baldwin}}, \bibnamefont{and}
	\bibinfo{author}{\bibfnamefont{A.}~\bibnamefont{Truscott}},
	\bibinfo{journal}{Nature} \textbf{\bibinfo{volume}{540}},
	\bibinfo{pages}{100} (\bibinfo{year}{2016}).
	
	\bibitem[{\citenamefont{Trimeche et~al.}(2020)\citenamefont{Trimeche, Lopez,
			Comparat, and Picard}}]{trimeche2020ion}
	\bibinfo{author}{\bibfnamefont{A.}~\bibnamefont{Trimeche}},
	\bibinfo{author}{\bibfnamefont{C.}~\bibnamefont{Lopez}},
	\bibinfo{author}{\bibfnamefont{D.}~\bibnamefont{Comparat}}, \bibnamefont{and}
	\bibinfo{author}{\bibfnamefont{Y.}~\bibnamefont{Picard}},
	\bibinfo{journal}{Physical Review Research} \textbf{\bibinfo{volume}{2}},
	\bibinfo{pages}{043295} (\bibinfo{year}{2020}).
	
	\bibitem[{\citenamefont{Wang et~al.}(2019)\citenamefont{Wang, Wang, Wang, Li,
			and Situ}}]{wang2019learning}
	\bibinfo{author}{\bibfnamefont{F.}~\bibnamefont{Wang}},
	\bibinfo{author}{\bibfnamefont{H.}~\bibnamefont{Wang}},
	\bibinfo{author}{\bibfnamefont{H.}~\bibnamefont{Wang}},
	\bibinfo{author}{\bibfnamefont{G.}~\bibnamefont{Li}}, \bibnamefont{and}
	\bibinfo{author}{\bibfnamefont{G.}~\bibnamefont{Situ}},
	\bibinfo{journal}{Optics Express} \textbf{\bibinfo{volume}{27}},
	\bibinfo{pages}{25560} (\bibinfo{year}{2019}).
	
	\bibitem[{\citenamefont{Rizvi et~al.}(2020)\citenamefont{Rizvi, Cao, Zhang, and
			Hao}}]{rizvi2020deepghost}
	\bibinfo{author}{\bibfnamefont{S.}~\bibnamefont{Rizvi}},
	\bibinfo{author}{\bibfnamefont{J.}~\bibnamefont{Cao}},
	\bibinfo{author}{\bibfnamefont{K.}~\bibnamefont{Zhang}}, \bibnamefont{and}
	\bibinfo{author}{\bibfnamefont{Q.}~\bibnamefont{Hao}},
	\bibinfo{journal}{Scientific Reports} \textbf{\bibinfo{volume}{10}},
	\bibinfo{pages}{1} (\bibinfo{year}{2020}).
	
	\bibitem[{\citenamefont{Wu et~al.}(2020)\citenamefont{Wu, Wang, Zhao, Xiao,
			Liang, Wang, Tian, Cheng, and Zhang}}]{wu2020deep}
	\bibinfo{author}{\bibfnamefont{H.}~\bibnamefont{Wu}},
	\bibinfo{author}{\bibfnamefont{R.}~\bibnamefont{Wang}},
	\bibinfo{author}{\bibfnamefont{G.}~\bibnamefont{Zhao}},
	\bibinfo{author}{\bibfnamefont{H.}~\bibnamefont{Xiao}},
	\bibinfo{author}{\bibfnamefont{J.}~\bibnamefont{Liang}},
	\bibinfo{author}{\bibfnamefont{D.}~\bibnamefont{Wang}},
	\bibinfo{author}{\bibfnamefont{X.}~\bibnamefont{Tian}},
	\bibinfo{author}{\bibfnamefont{L.}~\bibnamefont{Cheng}}, \bibnamefont{and}
	\bibinfo{author}{\bibfnamefont{X.}~\bibnamefont{Zhang}},
	\bibinfo{journal}{Optics and Lasers in Engineering}
	\textbf{\bibinfo{volume}{134}}, \bibinfo{pages}{106183}
	(\bibinfo{year}{2020}).
	
	\bibitem[{\citenamefont{Johnson et~al.}(2016)\citenamefont{Johnson, Alahi, and
			Fei-Fei}}]{johnson2016perceptual}
	\bibinfo{author}{\bibfnamefont{J.}~\bibnamefont{Johnson}},
	\bibinfo{author}{\bibfnamefont{A.}~\bibnamefont{Alahi}}, \bibnamefont{and}
	\bibinfo{author}{\bibfnamefont{L.}~\bibnamefont{Fei-Fei}}, in
	\emph{\bibinfo{booktitle}{European conference on computer vision}}
	(\bibinfo{organization}{Springer}, \bibinfo{year}{2016}), pp.
	\bibinfo{pages}{694--711}.
	
	\bibitem[{\citenamefont{Barbastathis et~al.}(2019)\citenamefont{Barbastathis,
			Ozcan, and Situ}}]{barbastathis2019use}
	\bibinfo{author}{\bibfnamefont{G.}~\bibnamefont{Barbastathis}},
	\bibinfo{author}{\bibfnamefont{A.}~\bibnamefont{Ozcan}}, \bibnamefont{and}
	\bibinfo{author}{\bibfnamefont{G.}~\bibnamefont{Situ}},
	\bibinfo{journal}{Optica} \textbf{\bibinfo{volume}{6}}, \bibinfo{pages}{921}
	(\bibinfo{year}{2019}).
	
	\bibitem[{\citenamefont{Lyu et~al.}(2017)\citenamefont{Lyu, Wang, Wang, Wang,
			Li, Chen, and Situ}}]{Lyu2017Deep}
	\bibinfo{author}{\bibfnamefont{M.}~\bibnamefont{Lyu}},
	\bibinfo{author}{\bibfnamefont{W.}~\bibnamefont{Wang}},
	\bibinfo{author}{\bibfnamefont{H.}~\bibnamefont{Wang}},
	\bibinfo{author}{\bibfnamefont{H.}~\bibnamefont{Wang}},
	\bibinfo{author}{\bibfnamefont{G.}~\bibnamefont{Li}},
	\bibinfo{author}{\bibfnamefont{N.}~\bibnamefont{Chen}}, \bibnamefont{and}
	\bibinfo{author}{\bibfnamefont{G.}~\bibnamefont{Situ}},
	\bibinfo{journal}{Scientific Reports} \textbf{\bibinfo{volume}{7}},
	\bibinfo{pages}{17865} (\bibinfo{year}{2017}).
	
	\bibitem[{\citenamefont{Song et~al.}(2022)\citenamefont{Song, Nie, Su, Chen,
			Zhou, Zhao, Peng, and Scully}}]{song20220}
	\bibinfo{author}{\bibfnamefont{H.}~\bibnamefont{Song}},
	\bibinfo{author}{\bibfnamefont{X.}~\bibnamefont{Nie}},
	\bibinfo{author}{\bibfnamefont{H.}~\bibnamefont{Su}},
	\bibinfo{author}{\bibfnamefont{H.}~\bibnamefont{Chen}},
	\bibinfo{author}{\bibfnamefont{Y.}~\bibnamefont{Zhou}},
	\bibinfo{author}{\bibfnamefont{X.}~\bibnamefont{Zhao}},
	\bibinfo{author}{\bibfnamefont{T.}~\bibnamefont{Peng}}, \bibnamefont{and}
	\bibinfo{author}{\bibfnamefont{M.~O.} \bibnamefont{Scully}},
	\bibinfo{journal}{Optics Communications} \textbf{\bibinfo{volume}{520}},
	\bibinfo{pages}{128450} (\bibinfo{year}{2022}).
	
	\bibitem[{\citenamefont{Collobert and Weston}(2008)}]{collobert2008unified}
	\bibinfo{author}{\bibfnamefont{R.}~\bibnamefont{Collobert}} \bibnamefont{and}
	\bibinfo{author}{\bibfnamefont{J.}~\bibnamefont{Weston}}, in
	\emph{\bibinfo{booktitle}{Proceedings of the 25th international conference on
			Machine learning}} (\bibinfo{year}{2008}), pp. \bibinfo{pages}{160--167}.
	
	\bibitem[{\citenamefont{Vaswani et~al.}(2017)\citenamefont{Vaswani, Shazeer,
			Parmar, Uszkoreit, Jones, Gomez, Kaiser, and
			Polosukhin}}]{vaswani2017attention}
	\bibinfo{author}{\bibfnamefont{A.}~\bibnamefont{Vaswani}},
	\bibinfo{author}{\bibfnamefont{N.}~\bibnamefont{Shazeer}},
	\bibinfo{author}{\bibfnamefont{N.}~\bibnamefont{Parmar}},
	\bibinfo{author}{\bibfnamefont{J.}~\bibnamefont{Uszkoreit}},
	\bibinfo{author}{\bibfnamefont{L.}~\bibnamefont{Jones}},
	\bibinfo{author}{\bibfnamefont{A.~N.} \bibnamefont{Gomez}},
	\bibinfo{author}{\bibfnamefont{{\L}.}~\bibnamefont{Kaiser}},
	\bibnamefont{and}
	\bibinfo{author}{\bibfnamefont{I.}~\bibnamefont{Polosukhin}},
	\bibinfo{journal}{Advances in neural information processing systems}
	\textbf{\bibinfo{volume}{30}} (\bibinfo{year}{2017}).
	
	\bibitem[{\citenamefont{Wolf et~al.}(2020)\citenamefont{Wolf, Debut, Sanh,
			Chaumond, Delangue, Moi, Cistac, Rault, Louf, Funtowicz
			et~al.}}]{wolf2020transformers}
	\bibinfo{author}{\bibfnamefont{T.}~\bibnamefont{Wolf}},
	\bibinfo{author}{\bibfnamefont{L.}~\bibnamefont{Debut}},
	\bibinfo{author}{\bibfnamefont{V.}~\bibnamefont{Sanh}},
	\bibinfo{author}{\bibfnamefont{J.}~\bibnamefont{Chaumond}},
	\bibinfo{author}{\bibfnamefont{C.}~\bibnamefont{Delangue}},
	\bibinfo{author}{\bibfnamefont{A.}~\bibnamefont{Moi}},
	\bibinfo{author}{\bibfnamefont{P.}~\bibnamefont{Cistac}},
	\bibinfo{author}{\bibfnamefont{T.}~\bibnamefont{Rault}},
	\bibinfo{author}{\bibfnamefont{R.}~\bibnamefont{Louf}},
	\bibinfo{author}{\bibfnamefont{M.}~\bibnamefont{Funtowicz}},
	\bibnamefont{et~al.}, in \emph{\bibinfo{booktitle}{Proceedings of the 2020
			conference on empirical methods in natural language processing: system
			demonstrations}} (\bibinfo{year}{2020}), pp. \bibinfo{pages}{38--45}.
	
	\bibitem[{\citenamefont{Liu et~al.}(2021)\citenamefont{Liu, Lin, Cao, Hu, Wei,
			Zhang, Lin, and Guo}}]{liu2021swin}
	\bibinfo{author}{\bibfnamefont{Z.}~\bibnamefont{Liu}},
	\bibinfo{author}{\bibfnamefont{Y.}~\bibnamefont{Lin}},
	\bibinfo{author}{\bibfnamefont{Y.}~\bibnamefont{Cao}},
	\bibinfo{author}{\bibfnamefont{H.}~\bibnamefont{Hu}},
	\bibinfo{author}{\bibfnamefont{Y.}~\bibnamefont{Wei}},
	\bibinfo{author}{\bibfnamefont{Z.}~\bibnamefont{Zhang}},
	\bibinfo{author}{\bibfnamefont{S.}~\bibnamefont{Lin}}, \bibnamefont{and}
	\bibinfo{author}{\bibfnamefont{B.}~\bibnamefont{Guo}}, in
	\emph{\bibinfo{booktitle}{Proceedings of the IEEE/CVF International
			Conference on Computer Vision}} (\bibinfo{year}{2021}), pp.
	\bibinfo{pages}{10012--10022}.
	
	\bibitem[{\citenamefont{Chen et~al.}(2013)\citenamefont{Chen, Peng, and
			Shih}}]{chen2013100}
	\bibinfo{author}{\bibfnamefont{H.}~\bibnamefont{Chen}},
	\bibinfo{author}{\bibfnamefont{T.}~\bibnamefont{Peng}}, \bibnamefont{and}
	\bibinfo{author}{\bibfnamefont{Y.}~\bibnamefont{Shih}},
	\bibinfo{journal}{Physical Review A} \textbf{\bibinfo{volume}{88}},
	\bibinfo{pages}{023808} (\bibinfo{year}{2013}).
	
	\bibitem[{\citenamefont{Candes et~al.}(2006)\citenamefont{Candes, Romberg, and
			Tao}}]{candes2006stable}
	\bibinfo{author}{\bibfnamefont{E.~J.} \bibnamefont{Candes}},
	\bibinfo{author}{\bibfnamefont{J.~K.} \bibnamefont{Romberg}},
	\bibnamefont{and} \bibinfo{author}{\bibfnamefont{T.}~\bibnamefont{Tao}},
	\bibinfo{journal}{Communications on Pure and Applied Mathematics: A Journal
		Issued by the Courant Institute of Mathematical Sciences}
	\textbf{\bibinfo{volume}{59}}, \bibinfo{pages}{1207} (\bibinfo{year}{2006}).
	
	\bibitem[{\citenamefont{Shi et~al.}(2017)\citenamefont{Shi, Jiang, Zhang, and
			Zhao}}]{shi2017deep}
	\bibinfo{author}{\bibfnamefont{W.}~\bibnamefont{Shi}},
	\bibinfo{author}{\bibfnamefont{F.}~\bibnamefont{Jiang}},
	\bibinfo{author}{\bibfnamefont{S.}~\bibnamefont{Zhang}}, \bibnamefont{and}
	\bibinfo{author}{\bibfnamefont{D.}~\bibnamefont{Zhao}}, in
	\emph{\bibinfo{booktitle}{2017 IEEE International Conference on Multimedia
			and Expo (ICME)}} (\bibinfo{organization}{IEEE}, \bibinfo{year}{2017}), pp.
	\bibinfo{pages}{877--882}.
	
	\bibitem[{\citenamefont{Zhang and Ghanem}(2018)}]{zhang2018ista}
	\bibinfo{author}{\bibfnamefont{J.}~\bibnamefont{Zhang}} \bibnamefont{and}
	\bibinfo{author}{\bibfnamefont{B.}~\bibnamefont{Ghanem}}, in
	\emph{\bibinfo{booktitle}{Proceedings of the IEEE conference on computer
			vision and pattern recognition}} (\bibinfo{year}{2018}), pp.
	\bibinfo{pages}{1828--1837}.
	
	\bibitem[{\citenamefont{Jiao et~al.}(2020)\citenamefont{Jiao, Gao, Feng, Lei,
			and Yuan}}]{jiao2020does}
	\bibinfo{author}{\bibfnamefont{S.}~\bibnamefont{Jiao}},
	\bibinfo{author}{\bibfnamefont{Y.}~\bibnamefont{Gao}},
	\bibinfo{author}{\bibfnamefont{J.}~\bibnamefont{Feng}},
	\bibinfo{author}{\bibfnamefont{T.}~\bibnamefont{Lei}}, \bibnamefont{and}
	\bibinfo{author}{\bibfnamefont{X.}~\bibnamefont{Yuan}},
	\bibinfo{journal}{Optics express} \textbf{\bibinfo{volume}{28}},
	\bibinfo{pages}{3717} (\bibinfo{year}{2020}).
	
	\bibitem[{\citenamefont{Goodfellow et~al.}(2016)\citenamefont{Goodfellow,
			Bengio, and Courville}}]{goodfellow2016deep}
	\bibinfo{author}{\bibfnamefont{I.}~\bibnamefont{Goodfellow}},
	\bibinfo{author}{\bibfnamefont{Y.}~\bibnamefont{Bengio}}, \bibnamefont{and}
	\bibinfo{author}{\bibfnamefont{A.}~\bibnamefont{Courville}},
	\emph{\bibinfo{title}{Deep learning}} (\bibinfo{publisher}{MIT press},
	\bibinfo{year}{2016}).
	
	\bibitem[{\citenamefont{Sutskever et~al.}(2014)\citenamefont{Sutskever,
			Vinyals, and Le}}]{sutskever2014sequence}
	\bibinfo{author}{\bibfnamefont{I.}~\bibnamefont{Sutskever}},
	\bibinfo{author}{\bibfnamefont{O.}~\bibnamefont{Vinyals}}, \bibnamefont{and}
	\bibinfo{author}{\bibfnamefont{Q.~V.} \bibnamefont{Le}},
	\bibinfo{journal}{Advances in neural information processing systems}
	\textbf{\bibinfo{volume}{27}} (\bibinfo{year}{2014}).
	
	\bibitem[{\citenamefont{Cho et~al.}(2014)\citenamefont{Cho,
			Van~Merri{\"e}nboer, Gulcehre, Bahdanau, Bougares, Schwenk, and
			Bengio}}]{cho2014learning}
	\bibinfo{author}{\bibfnamefont{K.}~\bibnamefont{Cho}},
	\bibinfo{author}{\bibfnamefont{B.}~\bibnamefont{Van~Merri{\"e}nboer}},
	\bibinfo{author}{\bibfnamefont{C.}~\bibnamefont{Gulcehre}},
	\bibinfo{author}{\bibfnamefont{D.}~\bibnamefont{Bahdanau}},
	\bibinfo{author}{\bibfnamefont{F.}~\bibnamefont{Bougares}},
	\bibinfo{author}{\bibfnamefont{H.}~\bibnamefont{Schwenk}}, \bibnamefont{and}
	\bibinfo{author}{\bibfnamefont{Y.}~\bibnamefont{Bengio}},
	\bibinfo{journal}{arXiv preprint arXiv:1406.1078}  (\bibinfo{year}{2014}).
	
	\bibitem[{\citenamefont{Schmidhuber et~al.}(1997)\citenamefont{Schmidhuber,
			Hochreiter et~al.}}]{schmidhuber1997long}
	\bibinfo{author}{\bibfnamefont{J.}~\bibnamefont{Schmidhuber}},
	\bibinfo{author}{\bibfnamefont{S.}~\bibnamefont{Hochreiter}},
	\bibnamefont{et~al.}, \bibinfo{journal}{Neural Comput}
	\textbf{\bibinfo{volume}{9}}, \bibinfo{pages}{1735} (\bibinfo{year}{1997}).
	
	\bibitem[{\citenamefont{Wang et~al.}(2004)\citenamefont{Wang, Bovik, Sheikh,
			and Simoncelli}}]{wang2004image}
	\bibinfo{author}{\bibfnamefont{Z.}~\bibnamefont{Wang}},
	\bibinfo{author}{\bibfnamefont{A.~C.} \bibnamefont{Bovik}},
	\bibinfo{author}{\bibfnamefont{H.~R.} \bibnamefont{Sheikh}},
	\bibnamefont{and} \bibinfo{author}{\bibfnamefont{E.~P.}
		\bibnamefont{Simoncelli}}, \bibinfo{journal}{IEEE transactions on image
		processing} \textbf{\bibinfo{volume}{13}}, \bibinfo{pages}{600}
	(\bibinfo{year}{2004}).
	
	\bibitem[{\citenamefont{Neimark et~al.}(2021)\citenamefont{Neimark, Bar, Zohar,
			and Asselmann}}]{neimark2021video}
	\bibinfo{author}{\bibfnamefont{D.}~\bibnamefont{Neimark}},
	\bibinfo{author}{\bibfnamefont{O.}~\bibnamefont{Bar}},
	\bibinfo{author}{\bibfnamefont{M.}~\bibnamefont{Zohar}}, \bibnamefont{and}
	\bibinfo{author}{\bibfnamefont{D.}~\bibnamefont{Asselmann}}, in
	\emph{\bibinfo{booktitle}{Proceedings of the IEEE/CVF International
			Conference on Computer Vision}} (\bibinfo{year}{2021}), pp.
	\bibinfo{pages}{3163--3172}.
	
	\bibitem[{\citenamefont{Nie et~al.}(2021{\natexlab{b}})\citenamefont{Nie, Song,
			Ren, Zhao, Zhang, Peng, and Scully}}]{nie2021deep}
	\bibinfo{author}{\bibfnamefont{X.}~\bibnamefont{Nie}},
	\bibinfo{author}{\bibfnamefont{H.}~\bibnamefont{Song}},
	\bibinfo{author}{\bibfnamefont{W.}~\bibnamefont{Ren}},
	\bibinfo{author}{\bibfnamefont{X.}~\bibnamefont{Zhao}},
	\bibinfo{author}{\bibfnamefont{Z.}~\bibnamefont{Zhang}},
	\bibinfo{author}{\bibfnamefont{T.}~\bibnamefont{Peng}}, \bibnamefont{and}
	\bibinfo{author}{\bibfnamefont{M.~O.} \bibnamefont{Scully}},
	\bibinfo{journal}{arXiv preprint arXiv:2112.13293}
	(\bibinfo{year}{2021}{\natexlab{b}}).
	
\end{thebibliography}
\end{document}